\documentclass[11pt]{article}
\usepackage{amssymb}
\usepackage{amsmath}
\usepackage{graphicx}
\usepackage{placeins}
\usepackage{booktabs}
\usepackage{multirow}
\usepackage{colortbl}
\usepackage{tabulary}
\usepackage{makecell}
\usepackage{tikz}
\usetikzlibrary{shapes,arrows.meta,positioning,backgrounds}
\usetikzlibrary{arrows.meta}
\usepackage[margin=1in]{geometry}
\usepackage[colorlinks=true,linkcolor=blue, citecolor=blue]{hyperref}
\usepackage[style=authoryear,natbib]{biblatex}
\title{(Anti)Gravitron: A Statistical Physics Perspective on Multidimensional Metrics of Polarizing Inequality}
\author{Rolando Gonzales Martinez$^{1,2}$}
 \date{%
     $^1$Rijksuniversiteit Groningen (Kingdom of the Netherlands) \\%
     $^2$University of Oxford (United Kingdom) \newline contact: \textcolor{blue}{r.m.gonzales.martinez@rug.nl} \\[2ex]%
     \today
 }

\begin{document}

\maketitle

\begin{abstract}
\noindent 
This paper introduces a novel framework for measuring multidimensional inequality based on a statistical physics reinterpretation of centrifugal and centripetal forces in rotating systems. Inspired by the mechanics of the Gravitron and extended via the conceptual AntiGravitron---a thought experiment that models the polarization dynamics of social inclusion and exclusion---this study proposes a new class of inequality metrics grounded in multivariate mixtures of Beta distributions. These composite metrics capture three key structural dimensions of polarizing inequality: the number and balance of population clusters (modal entropy), the internal uniformity of each cluster (concentration), and the separation between clusters in attribute space (geometric divergence). Building on analogies from magnetohydrodynamics, non-equilibrium statistical mechanics, and entropy-based modeling, the AntiGravitron conceptualizes inequality not as a scalar deviation from equality, but as an emergent property of multimodal distributional geometry in bounded domains. Monte Carlo simulations of the AntiGravitron show how bifurcation, stratification, and polarization jointly influence inequality measurements due to multidimensional attraction-repulsion forces. An empirical application of the AntiGravitron to US household income data reveals polarized inequality driven by intersecting centripetal and centrifugal socio-economic forces affecting Black and African American populations in the Philadelphia County. By bridging physical systems theory and social stratification analysis, this paper offers a rigorous, flexible, and interpretable metric that enhances the understanding of polarizing inequality in high-dimensional, structurally heterogeneous contexts. The AntiGravitron framework holds promise for small-area estimation of polarization-driven inequality in socio-economics and biomedicine domains such as epidemiology, where inequality arises from multi-axial exclusion and attractor dynamics.
\end{abstract}



\newpage

\section{Introduction}

Inequality is a ubiquitous feature of complex systems, emerging not merely as a socio-economic artifact but as a statistical regularity across domains as diverse as epidemiology, information theory, fluid dynamics, and wealth distribution. Traditional scalar metrics such as the Gini, Theil, and Atkinson indices reduce this complexity into one-dimensional aggregates, yet fail to capture the full geometry of multimodal, bounded, and stratified systems. This paper proposes a novel multidimensional inequality metric inspired by the non-nomological AntiGravitronian potential, derived from inverting the statistical mechanics of centrifugal and centripetal forces observed in physical systems such as the Gravitron.

Centrifugal and centripetal forces arise naturally when analyzing motion in rotating reference frames. In such non-inertial systems, objects experience two inertial effects: a velocity-dependent Coriolis force and a radial centrifugal force pointing outward from the center of rotation. The apparent motion is stabilized by a real, physical centripetal force—supplied by tension, gravity, friction, or electromagnetic interaction—that pulls objects inward along a curved path. In centrifuges, spinning platforms push dense particles outward to separate components; in particle accelerators, magnetic fields provide centripetal forces that bend charged particles into controlled orbits; in Gravitron rides, individuals are pinned to the wall by intense centrifugal acceleration as the structure spins rapidly, while centripetal normal forces from the rotating drum maintain confinement.

This paper extends the physical principles of the Gravitron into a new socio-physical framework by proposing the \textit{AntiGravitron}---a conceptual device introduced as a \textit{Gedankenexperiment}---that reverses the classical centrifugal and centripetal roles observed in physical rotation. In the AntiGravitron machine, inequality emerges not from equilibrium rotation but from a structural imbalance of social forces: centripetal elitist forces attract and retain insiders toward central positions within homophily networks (social, economic, or institutional), while centrifugal exclusionary forces expel others toward the periphery, thus generating a polarization phenomenon. The centrifugal and magnetic forces---economic, institutional, political, spatial, cultural, and technological---interact dynamically to produce multidimensional patterns of exclusion and segregation. The result is a bifurcated system where agents are clustered into distinct basins of attraction, generating multimodal probability distributions over multidimensional attribute spaces. Unlike the Gravitron, where all agents are subject to uniform centrifugal forces, in the AntiGravitron inequality is structurally stratified and polarized: centripetal forces concentrate social advantage, while centrifugal forces diversify and fragment the marginalized. From a statistical physics perspective, this landscape is analogous to a rotating ensemble with competing attractors, where the effective potential is asymmetric and dynamically evolving.

The resulting distribution of particles obtained from the AntiGravitron was modeled with multivariate mixtures of Beta distributions that capture polarizing inequality. These distributions are naturally bounded on $[0,1]^K$, and are capable of modeling multimodality, heavy tails, and concentration within and across modes, thus enabling the quantification of polarizing inequality in systems characterized by heterogeneity, stratification, and emergent asymmetries. The metric of polarizing inequality proposed in this study is a synthesizes of three dimensions: modal entropy (fragmentation), intra-modal concentration (homogeneity), and inter-modal distance (polarization). Through Monte Carlo simulations and the empirical analysis of U.S. income disparities by ethnicity and geography, this paper shows how the AntiGravitron provides a mathematically grounded tool for capturing the emergent geometry of polarizing inequality in complex adaptive systems. The data and Python scripts that can be used to reproduce the results of this study are available in the \href{https://github.com/rogon666/AntiGravitron}{AntiGravitron GitHub repository}.

\section{Background on inequality measures}

Inequality is a common feature of complex, self-organizing systems in economics, health, or physics, where outcomes are highly skewed due to a small fraction of agents, events, or entities accounting for a disproportionately large share of the total. Such heavy-tailed distributions emerge without central design, hinting at underlying processes of self-organization and possibly critical dynamics.  

Key inequality metrics are the Gini, Theil, and Atkinson indices. The Gini index is a summary statistic that measures how equitably a resource is distributed in a population \citep{farris2010gini}. It gauges the concentration of a distribution by comparing cumulative proportions of population and income (graphically, it is twice the area between the Lorenz curve and the diagonal of equality). Gini ranges from 0 (perfect equality) to 1 (extreme inequality where one unit holds all resources). Statistically, it can be expressed as half the relative mean absolute difference of incomes. The Gini is an anonymous measure (ignoring individual identities) and satisfies the Pigou–Dalton transfer principle (meaning any progressive transfer reduces Gini). However, Gini has well-known limitations in terms of statistical properties \citep{osberg2017limitations}. First, it only provides a partial ordering of distributions: if Lorenz curves cross, Gini alone cannot tell which distribution is more unequal. Second, the Gini is not decomposable into within- and between-group components; unlike some measures, it cannot cleanly attribute inequality to subpopulation contributions. Third, it is weakly transfer-sensitive, meaning that its response to a given income transfer depends only on the rank positions of donor and recipient, not on the income levels; transfers at different parts of the distribution can have similar impact if ranks are similar. In practice, this implies Gini is most sensitive to transfers around the middle of the distribution and less sensitive to changes at the extremes. 

The Theil index is an entropy-based inequality measure belonging to the family of generalized entropy indices \citep{maria2017theil}. It is defined (in its basic form) as $T = \frac{1}{n}\sum_i \frac{y_i}{\bar{y}}\log\frac{y_i}{\bar{y}}$, which intuitively measures the divergence of the income distribution from uniformity (each term compares individual income $y_i$ to the mean $\bar{y}$, weighted by income share). By dividing by the maximum possible value (which for $T$ is $\log n$ when one of $n$ individuals has all income), the Relative Theil is scaled to range between 0 and 1. This scaling yields an interpretable 0-1 index where 1 corresponds to the most unequal state (one person holds all income) and 0 to perfect equality. A key advantage of the Theil index is additive decomposability: total inequality can be exactly decomposed into between-group and within-group components for any partition of the population. This property is extremely useful for studying spatial or group disparities. Additionally, the Theil index is strongly transfer-sensitive, meaning it reacts more to income transfers involving the poor versus rich (or vice versa) than to transfers among equals. In other words, a transfer of a given amount from a rich person to a poor person will reduce Theil index more than the same transfer between two high-income persons, reflecting a stronger concern for transfers at the tails. Because of its roots in information theory, the Theil index can also be interpreted as the entropy (or redundancy) of the income distribution.

The Atkinson index \citep{atkinson1970measurement}, takes a distinctly normative approach to inequality measurement. It explicitly incorporates a social welfare function and a value judgment about inequality aversion via a parameter $\epsilon$. The Atkinson index is defined as $A(\epsilon) = 1 - \frac{\tilde{\mu}}{\bar{\mu}}$, where $\bar{\mu}$ is the actual mean income and $\tilde{\mu}$ is the equally distributed equivalent income, i.e. the level of income per person that would give the same social welfare as the current distribution (under a specified utility function). If incomes were perfectly equal, everyone would have $\bar{\mu}$ and welfare would be maximal; with inequality, $\tilde{\mu} < \bar{\mu}$, and the Atkinson index captures that proportional shortfall. The parameter $\epsilon \ge 0$ governs the degree of inequality aversion: higher $\epsilon$ means society (or the analyst) is more sensitive to inequality at the lower end (giving extra weight to poorer individuals in the social welfare function), whereas $\epsilon$ close to 0 implies near-indifference to inequality (focus on efficiency). In practical terms, with $\epsilon$ larger, the Atkinson index will rise (show more inequality) if the lower tail gets worse, reflecting that such deterioration has a big welfare cost. With $\epsilon$ very small, Atkinson’s index approaches zero even for unequal distributions. Atkinson’s family of measures has several attractive properties: it is sensitive to different parts of the distribution depending on $\epsilon$, it satisfies the Pigou–Dalton principle for all $\epsilon > 0$, and interestingly it is multiplicatively decomposable (a less common form of decomposition) under certain conditions. It also provides a complete ranking of distributions when $\epsilon$ is fixed, meaning any two distributions can be welfare-ranked (unlike Gini, which fails if Lorenz curves cross). 

Inequality across disciplines reflects universal heavy-tailed patterns. One of the remarkable insights from complexity science and socio-economic analysis is that skewed, heavy-tailed distributions--where a few entities have enormously more than the rest--appear in many domains. The classic 80/20 Pareto principle \citep{blanchet2022generalized} was first noted in wealth: roughly 20\% of people hold 80\% of the wealth. But analogous patterns surface in diverse fields, underscoring a universal inequality phenomenon. The distribution of scientific citations across papers or authors is highly unequal, for example, often following a Lotka’s law or power-law form \citep{saam1999lotka}. A small number of papers accumulate a huge number of citations, while the majority receive few, a pattern mirroring income/wealth concentration \citep{aytac2025lotka}. The Hirsch index (h-index) was proposed as an individual-level metric to summarize a scientist’s citation profile \citep{bertoli2017theoretical}. While not an inequality measure per se, the h-index attempts to balance productivity and impact by finding the largest number $h$ such that the scientist has $h$ papers with at least $h$ citations each. In doing so, it implicitly downplays the effect of a single outlier paper: no matter how extremely cited one paper is, it can raise $h$ by at most 1 once it far exceeds $h$ citations. This makes the h-index less sensitive to the extreme tail than, say, total citation count (analogous to how Atkinson or Theil can be tuned to emphasize different parts of the distribution). Nonetheless, the existence of extremely highly cited papers (“citation blockbusters”) and the steep drop-off for the rest means the citation distribution has a high Gini coefficient, since  a few publications or researchers garner a large share of all citations \citep{bartneck2011detecting}. In bibliometrics, researchers sometimes compute inequality measures on citation distributions: for example, a Gini coefficient of citations can quantify how concentrated a journal’s citations are among its papers, or how unequal citations are across researchers in a field. These analogues highlight the universality of inequality: the same statistical tools (Lorenz curves, tail exponent estimates, etc.) applied to incomes can describe the scientific fame or impact distribution.

Public health studies have revealed also that not all infectious individuals contribute equally to an outbreak. In diseases like SARS or COVID-19, a minority of superspreaders infect far more secondary contacts than the typical patient. This is quantified by the famous 20/80 rule in epidemiology: approximately 20\% of infected persons can be responsible for about 80\% of transmissions \citep{cheng2025power}. Such a distribution of secondary cases per index case is extremely skewed, most people transmit to few or no others, while a few unlucky or high-contact individuals cause dozens of new cases. Statistically, this is often modeled with a heavy-tailed or over-dispersed distribution (like a negative binomial with small dispersion parameter $k$). The inequality of infection counts can be interpreted through the same lens as income inequality. For instance, one could calculate a Gini index for the distribution of secondary infections: it would be very high (close to 1 in extreme superspreading scenarios) reflecting that a large share of infections is caused by a tiny fraction of infectors. Indeed, epidemiologists have introduced measures that capture the fraction of transmissions attributable to the most infectious people \citep{peter2022fractional}, which directly parallels the concept of “top income share” in economics. The fractional transmission in an epidemic is analogous to saying the top quintile holds of income: it signals a massive imbalance. The implications of this inequality are practical: targeting interventions (vaccines, contact reduction) to the potential superspreaders could dramatically reduce total spread. In terms of self-organizing systems, disease transmission networks self-organize such that a few nodes (individuals) become hubs of infection. 

In statistical physics and complexity science, power-law distributions are hallmarks of phenomena like earthquakes (frequency vs magnitude), avalanches in sandpile models, city sizes, and financial asset fluctuations \citep{sornette2009seismicity, boccara2010power}. These systems often exhibit self-organized criticality, meaning they naturally evolve to states where no single scale dominates event sizes, leading to many small events and a few giant ones. The result is a heavy tail, indicating inequality in event sizes. For example, city population distributions follow Zipf’s law, so the largest city in a country can be multiple times bigger than the second-largest, and so on \citep{arshad2018zipf}. In a self-organizing view, this may result from cumulative advantage processes, analogous to processes that yield income or citation inequality. In networks, preferential attachment models produce a few nodes with extremely high degree (connections) while most nodes have few, a degree inequality that mirrors wealth inequality \citep{jeong2003measuring}. Newman and others have noted that power-law tails (Pareto distributions) in wealth or firm sizes are not unique to economics but appear in many complex systems, often linked to multiplicative stochastic processes. The distribution of wealth in an economy has been likened to the distribution of energy among particles in a gas, but with an added mechanism for a heavy tail, e.g., allowing some particles to gain repeatedly \citep{modanese2016common}. 

In biomedicine and public health, heavy-tailed distributions appear, for example, in healthcare spending: a small fraction of patients (often those with complex chronic conditions) account for the bulk of healthcare expenditures \citep{berk2001concentration}. This reflects an inequality in health resource utilization and needs, sometimes called the problem of super-users in healthcare \citep{yuan2015mixed}. In genetics, few hub genes participate in many interactions while most have few, showing again a network inequality \citep{boucher2013genetic}. Each of these cases can be analyzed with inequality metrics akin to those used for income. Despite the different mechanics of these systems, their outcome distributions share a common skewed form, and often similar mathematical models (power laws, log-normal mixtures, etc.) describe them. This universality suggests that inequality emerges from a combination of random processes, feedback loops, and constraints that are common across systems.

There is also a relationship between inequality and the concept of polarization. Polarization is the process in which a distribution splits into distinct, internally homogeneous and externally heterogeneous groups, such that within-group identification and between-group alienation intensify, which leads to social tensions and potential conflict \citep{esteban1994polarization}. \citet{stewart2020polarization} presented a formal model of polarization based on cultural evolution and evolutionary game theory. They showed that economic decline and rising inequality incentivize individuals to adopt risk-averse, in-group-favoring strategies to mitigate personal economic risk. These strategies limit cross-group interaction and propagate through social learning. Even after the initial economic conditions reverse, polarization can persist due to path dependency. 

\citet{duclos2004polarization} developed an axiomatic framework to measure polarization based on the identification-alienation model. In this framework, polarization increases when individuals identify more strongly with their own group (identification) and feel alienated from others (alienation). \citet{duclos2004polarization} derive a class of polarization indices satisfying this structure and provide methods for consistent statistical estimation. They show that while inequality measures general dispersion, polarization captures the emergence of antagonistic and cohesive subgroups within a distribution. \citet{gregg2008two} also investigate polarization in the labor market by comparing individual and household-level joblessness in Britain. While the individual non-employment rate remained relatively constant over decades, the household-based joblessness rate rose, indicating a polarization in the distribution of work. The increase in households where all adults are either fully employed or fully unemployed, and the decline of mixed-status households, signifies rising within-household inequality and structural segmentation of labor outcomes. Taken together, these studies converge on the idea that polarization is a distinct, multidimensional phenomenon that can emerge from inequality but is not reducible to it. While inequality measures dispersion (e.g., via the Gini coefficient), polarization focuses on the formation of internal clusters with heightened internal cohesion and external antagonism. 

Economically, inequality can lead to risk-averse social behavior, reinforcing in-group preferences and undermining cooperative norms. Statistically, this divergence may manifest in multimodal income or employment distributions, and behaviorally, it expresses itself through increasing partisanship, social distrust, or labor market bifurcation. Thus, theoretical and empirical models of polarizing inequality need to take into account that polarization is not only a consequence of inequality but also a mechanism that sustains and amplifies it. Inequality indices like Gini, Theil (relative), and Atkinson are all bounded between 0 and 1 (for the definitions used), but empirically, when calculated from sample survey data, their sampling distributions tend to be skewed and heavy-tailed. For example, if a small area has very few sampled households, the computed Gini could by chance be extremely high or low (heavy tails), and the distribution is not symmetric (e.g., Gini is bounded below by 0 but can bunch up near higher values if one sampled household is much richer than others). 

Recent methodological innovation have proposed  Beta distributions for the estimation of  multimodal inequality. \citet{von2016robust} introduced a multimodel generalized‑Beta estimator for binned-income data, demonstrating extended Beta distribution families’ utility in inequality estimation. Traditional methods, like the Fay-Herriot model, assume a normal (Gaussian) sampling distribution for the direct estimates. That assumption is inadequate for skewed, bounded metrics. An alternative is to use distributions naturally defined on \([0,1]\), like the Beta distribution, which is often used for proportions or rates. However, a single Beta distribution may still be too restrictive when the data exhibit marked skewness or multi-modality. The Flexible Beta model introduced by \citet{de2024small} addresses this skewness and multi-modality by using a mixture of two Beta distributions. In essence, it allows the area-level random effect (or the sampling distribution of the estimator) to be a two-component mixture, which greatly increases flexibility. A two-Beta mixture has four shape parameters (two for each component) and a mixing weight, enabling it to mimic a wide variety of distribution shapes, from strongly skewed to approximately symmetric, uni-modal or bi-modal, and with potential heavy tails in the sense of extra mass near 0 or 1 beyond what a single Beta could capture. This flexibility is crucial because the inequality estimates across areas might themselves have an uneven distribution, for example, most regions might cluster around a moderate Gini value \~0.3–0.4, but a few regions with very diverse economies could have Gini above 0.5 (forming a heavy upper tail). The mixture model can accommodate such a scenario, essentially by allocating a small proportion of areas to a high-inequality Beta component (capturing the tail) while the bulk of areas follow another Beta distribution centered at the moderate inequality level. This approach is well-aligned with the theme of universality in inequality: it embraces the idea that the distribution of inequality itself might be heavy-tailed (some areas much more unequal than others, just as some individuals are much richer than others) and uses a flexible, heavy-tail-capable model to capture it. 

From a self-organization perspective, inequality itself might be a sign of a system organizing into a critical or near-critical state. In financial markets, for example, a build-up of wealth in a few hands can increase systemic fragility. In social systems, extreme inequality can indicate positive feedback loops that have run unchecked (e.g., monopolies accumulating power, or superstar academics getting disproportionate attention). These processes are emergent in that they are not centrally imposed but arise from many local interactions, yet they produce a consistent global pattern (the heavy tail). Mixtures of Betas capture complex (multi-modal) data structures by combining multiple components, each with its own shape and weight. This flexibility enables the representation of heterogeneous subpopulations, which is often a marker of inequality. In small-area estimation of income inequality, for example, previous research \citep{de2024small} found that Bayesian hierarchical models using mixtures of Betas outperformed standard Beta regression, achieving lower bias and better coverage at regional levels when compared to traditional metrics of inequality like the Gini index and the Atkinson coefficient.

\section{The (Anti)Gravitron}

\begin{figure}
    \centering
    \includegraphics[width=0.75\linewidth]{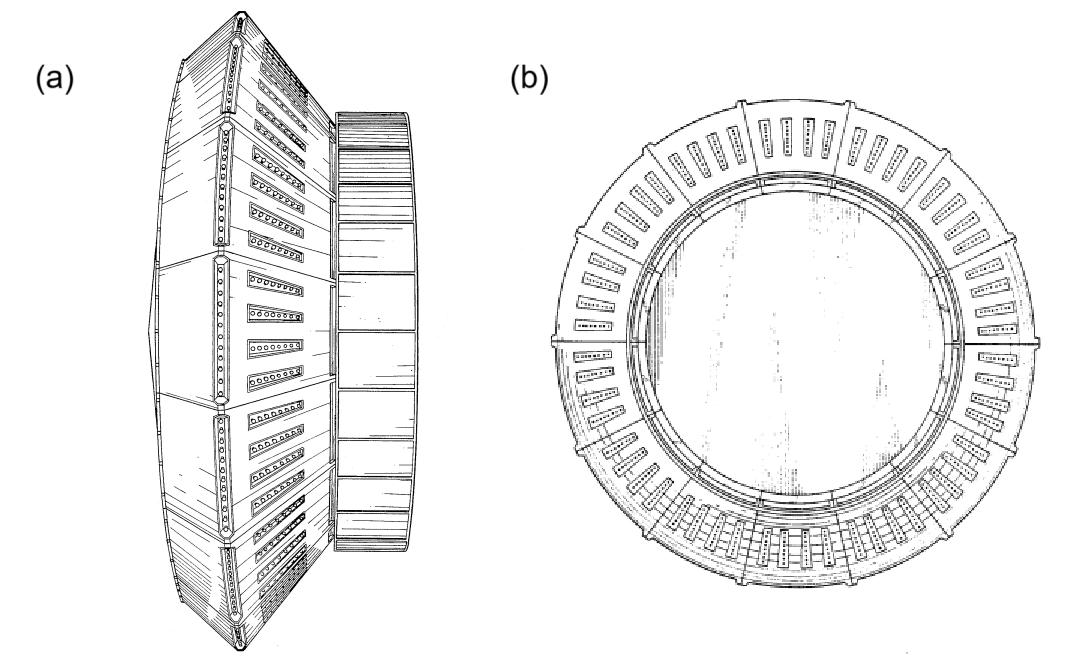}
    \caption{Side elevation (a) and plan view (b) of the Gravitron centrifugal machine, showing the inward-sloping support panels, drive housing ring, and circular floor assembly. Source: U.S. Design Patent USD284601S.}
    \label{fig:gravitron}
\end{figure}
\subsection{The Gravitron}

The Gravitron, first introduced in 1983 as an evolution of the mid-century Rotor. It employs a circular, enclosed chamber lined with forty-eight evenly spaced, inward-sloping support panels that, when spun to approximately 24 RPM by a 33 kW motor, generate centrifugal forces up to three times Earth’s gravity, allowing riders to remain suspended when the floor retracts. As shown in Figure \ref{fig:gravitron}, the front elevation has an outer shell composed of vertically ribbed cladding panels and an overhanging roof ring that conceals the drive mechanism, while a narrow trumpet-shaped entrance aperture aligns with a removable safety gate. From the top view Figure \ref{fig:gravitron}a, a star-burst arrangement of radial support arms connects the central hub to the peripheral ring, forming a rigid, symmetrical lattice that ensures even load distribution and facilitates rapid assembly on portable platforms. The bottom plan (\ref{fig:gravitron}b) exposes a circular floor plate punctuated by retractable floor segments and peripheral drainage channels. From an engineering perspective, the Gravitron exemplifies the application of basic centrifugal principles for generating immersive simulation of elevated gravity.

\subsection{The AntiGravitron}

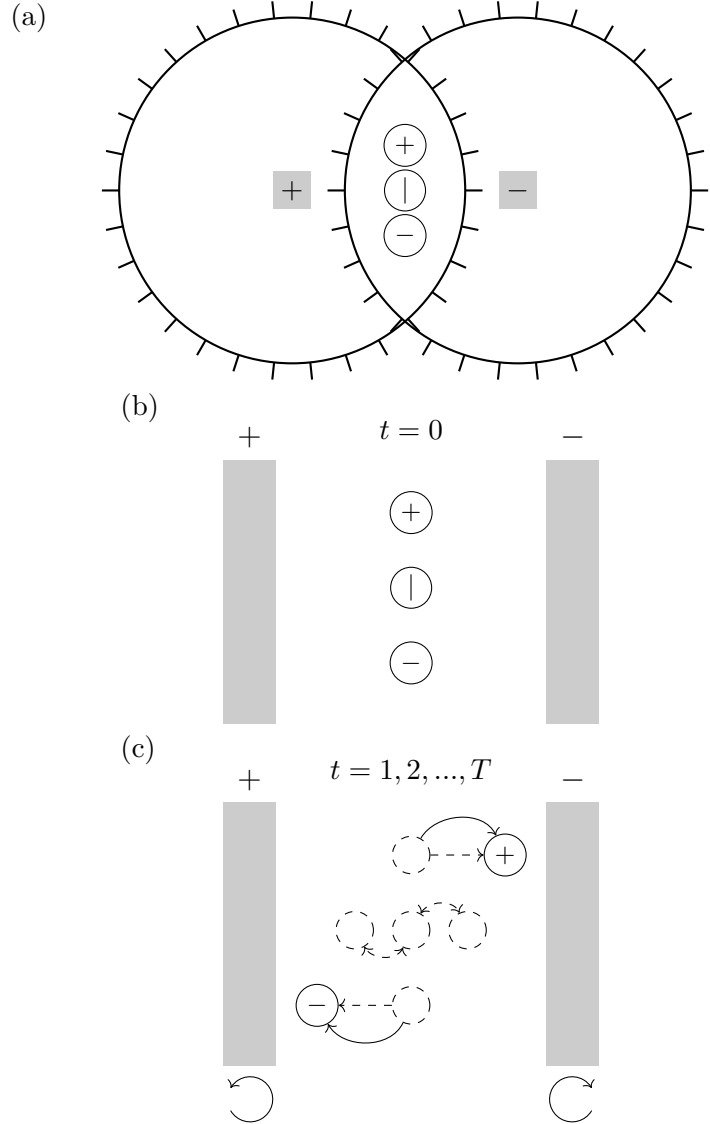
\begin{figure}[ht!]
\centering
\begin{tikzpicture}
\node at (2.5,2.3) {(a)};
\def\radius{2.3}
\def\n{30} 

\draw[thick] (6,0) circle (\radius);
\draw[thick] (9,0) circle (\radius);

\foreach \i in {0,12,...,348} {
  \draw[thick] 
    ({6 + \radius*cos(\i)}, {\radius*sin(\i)}) -- 
    ({6 + 1.1*\radius*cos(\i)}, {1.1*\radius*sin(\i)});
}

\foreach \i in {0,12,...,348} {
  \draw[thick] 
    ({9 + \radius*cos(\i)}, {\radius*sin(\i)}) -- 
    ({9 + 1.1*\radius*cos(\i)}, {1.1*\radius*sin(\i)});
}

\filldraw[fill=gray!40, draw=none] (6,0) ++(-0.25,-0.25) rectangle ++(0.5,0.5);
\filldraw[fill=gray!40, draw=none] (9,0) ++(-0.25,-0.25) rectangle ++(0.5,0.5);

\node at (6,0) { $+$};
\node at (9,0) { $-$};

\node[draw, circle, inner sep=2pt] at (7.5, 0.6) {\small $+$};
\node[draw, circle, inner sep=2pt] at (7.5, 0.0) {\small $|$};
\node[draw, circle, inner sep=2pt] at (7.5, -0.6) {\small $-$};
\end{tikzpicture}

\begin{tikzpicture}
\node[anchor=west] at (-4,4.2) {(b)};
\node at (0,3.9) {$t = 0$};
\fill[gray!40] (-2.5,0) rectangle (-1.8,3.5);  
\fill[gray!40] (1.8,0) rectangle (2.5,3.5);    
\node at (-2.15,3.8) {\large $+$};
\node at (2.15,3.8) {\large $-$};
\node[draw, circle, inner sep=2pt] at (0,2.8) {\small $+$};
\node[draw, circle, inner sep=2pt] at (0,1.8) {\small $|$};
\node[draw, circle, inner sep=2pt] at (0,0.8) {\small $-$};
\end{tikzpicture}

\begin{tikzpicture}
\node[anchor=west] at (-4,4.2) {(c)};
\node at (0,3.9) {$t = 1,2,...,T$};
\fill[gray!40] (-2.5,0) rectangle (-1.8,3.5);  
\fill[gray!40] (1.8,0) rectangle (2.5,3.5);    
\node at (-2.15,3.8) {\large $+$};
\node at (2.15,3.8) {\large $-$};
\node[draw, circle, inner sep=2pt] (c1) at ( 1.25,2.8) {\small $+$};
\node[draw, circle, inner sep=2pt] (c3) at (-1.25,0.8) {\small $-$};

\node[draw, dashed, circle, inner sep=5pt] (dc1) at ( 0,2.8) {};
\node[draw, dashed, circle, inner sep=5pt] (dcr) at ( 0.75,1.8) {};
\node[draw, dashed, circle, inner sep=5pt] (dc2) at ( 0,1.8) {};
\node[draw, dashed, circle, inner sep=5pt] (dcl) at (-0.75,1.8) {};
\node[draw, dashed, circle, inner sep=5pt] (dc3) at ( 0,0.8) {};

\draw[->] (-2.4,-0.6) arc[start angle=210, end angle=510, radius=0.3];
\draw[->, xscale=-1] (-2.4,-0.6) arc[start angle=210, end angle=510, radius=0.3];

\draw[dashed,->] (dc1) -- (c1);
\draw[dashed,->] (dc3) -- (c3);
\draw[->] (dc1) to[out=60, in=115] (c1);  
\draw[->] (dc3) to[out=-115, in=-60] (c3);
\draw[dashed, <->] (dc2) to[out=60, in=115] (dcr);
\draw[dashed, <->] (dc2) to[out=-115, in=-60] (dcl);
\end{tikzpicture}

   \caption{AntiGravitron dynamics with two counter-rotating rotors and three interacting particles. \textbf{(a)} Planar (top-down) schematic of the AntiGravitron device, composed of two overlapping circular rotors, each containing a central dynamo spindle generating magnetic fields of opposite polarity. The left spindle is positively charged, while the right spindle is negatively charged. Three particles—positive ($+$), neutral ($|$), and negative ($-$)—are symmetrically initialized within the overlapping region of the rotors. \textbf{(b)} Elevation view of the system at rest ($t = 0$), representing the initial state of reflectional symmetry and spatial equilibrium, where no net force induces charge-dependent displacement. \textbf{(c)} As the rotors initiate rotation at time $t = 1, 2, \dots, T$, dynamical asymmetry emerges due to the interaction between centrifugal forces and Lorentz magnetic forces. The positively charged particle migrates toward the negatively charged spindle (right), which rotates clockwise, while the negatively charged particle is attracted to the positively charged spindle (left), which rotates counterclockwise. Neutrally charged particles undergo inertial oscillations. Dashed circles illustrate intermediate positions and the solid curves indicate the non-linear trajectories caused by the Coriolis effect. This transition from symmetric to asymmetric charge distributions models a bifurcation process analogous to the emergence of inequality under competing attractors.}
    \label{fig:AntiGravitron}
\end{figure}

The \textit{AntiGravitron} is a conceptual device for modeling symmetry, bifurcation, and emergent polarizing inequality within a rotating field system that combines mechanical rotation with magnetodynamic forces. Inspired by the classical Gravitron yet fundamentally distinct, the AntiGravitron in its simplest design consists of two counter-rotating rings, each housing a dynamo spindle that generates magnetic fields of opposite polarity. At time $t = 0$ (Figures~\ref{fig:AntiGravitron}a and~\ref{fig:AntiGravitron}b), the system is in a state of symmetric equilibrium: three particles with charge $q \in \{-1, 0, +1\}$ are evenly distributed within the overlapping region of the rings. In this configuration, no net force induces spatial differentiation, and the expected radial displacement $\mathbb{E}[r_q]$ is identical for all $q$, preserving reflectional symmetry.

Once the rings begin to rotate at $t > 0$ (Figure~\ref{fig:AntiGravitron}c), the system transitions into a dynamically asymmetric regime. The motion of each particle is influenced by the combined effects of centrifugal force $\vec{F}_\mathrm{cent} = m\omega^2 \vec{r}$ and Lorentz force $\vec{F}_\mathrm{mag} = q\vec{v} \times \vec{B}$, where $\vec{B}$ denotes the magnetic field produced by the dynamo spindles \citep{moffatt1978magnetic}. The effective potential governing particle behavior can be written as
\[
V(q, \vec{r}) = \tfrac{1}{2} m \omega^2 r^2 - q \, \Phi(\vec{r}),
\]
where $\Phi(\vec{r})$ is the magnetic scalar potential with opposing signs across the two spindles. This interaction leads to spatial segregation by charge: positively charged particles are attracted to the negatively charged spindle and vice versa, while neutral particles undergo oscillatory inertial motion without magnetic attraction. As the system evolves, it undergoes spontaneous symmetry breaking \citep{anderson1972more}, and the particle distribution bifurcates into charge-dependent basins of attraction. This results in a transition from unimodal to bimodal distributions in physical space, analogous to the formation of inequality in a one-dimensional socioeconomic structure. The asymmetry is characterized by $\mathbb{E}[r_+] \neq \mathbb{E}[r_-]$ and an energy gap $\Delta E = |V_+ - V_-|$, both of which serve as physical proxies for measuring inequality in such systems.

In an AntiGravitron system with two rotating rings and a large number of positively and negatively charged particles, the resulting statistical distribution of particle positions becomes bimodal, skewed, and heavy-tailed. This complex spatial distribution can be effectively modeled using a mixture of Beta distributions. 
Extending the AntiGravitron analogy, one can consider a generalized \textit{multi-spindle} and \textit{multi-particle} AntiGravitron in which multiple dynamo spindles of varying strength and polarity are distributed across a system with a large population of particles. This extension corresponds to a society composed of multiple elite and non-elite subgroups, each exerting centripetal forces that retain insiders and centrifugal forces that repel outsiders, potentially shaped by Coriolis-type inertial dynamics \citep{vallis2017atmospheric}. From a statistical physics perspective, the resulting inequality landscape can be modeled by a multivariate multimodal distribution. Here, each dimension represents an axis of stratification (e.g., wealth, education, influence), while each mode (spindle) corresponds to a distinct social cluster or attractor basin. The interaction of these dimensions gives rise to complex patterns of polarization, network repulsion, and homophily, quantifiable through entropy-based or divergence-based multidimensional inequality metrics \citep{cowan1996nonequilibrium, jones2015econophysics}. The AntiGravitron, in this extended form, offers a rich \textit{gedankenexperiment} for theorizing inequality in polarizing systems where social and bio-medical forces jointly shape emergent structures of advantage and exclusion.

\section{Statistical mechanics of centrifugal and centripetal forces in discrete and continuous multivariate multimodal distributions}

\subsection{Centrifugal and centripetal forces} 

Statistical mechanics in a rotating frame is formulated by introducing an effective Hamiltonian 
\[
H_{\rm eff}(\mathbf{p},\mathbf{r}) = \frac{\mathbf{p}^2}{2m} + U(\mathbf{r}) - \boldsymbol{\omega}\mathbf{L}
\]
where \(\mathbf{p}\) is the particle momentum, \(m\) its mass, \(U(\mathbf{r})\) the conservative potential, \(\boldsymbol{\omega}\) the rotation vector, and \(\mathbf{L} = \mathbf{r}\times\mathbf{p}\) the angular momentum.  The term \(-\boldsymbol{\omega}\mathbf{L}\) generates both the centrifugal and Coriolis effects in the Boltzmann weight 
\[
f(\mathbf{p},\mathbf{r}) = \frac{1}{Z}\exp\bigl(-\beta H_{\rm eff}(\mathbf{p},\mathbf{r})\bigr)
\]
with inverse temperature \(\beta=1/(k_{\rm B}T)\), \(k_{\rm B}\) Boltzmann’s constant, and partition function
\[
Z = \frac{1}{h^{3N}N!}\!\int\!d^{3N}r\,d^{3N}p\;\exp\bigl(-\beta H_{\rm eff}\bigr).
\]

Projecting into cylindrical coordinates \((r,\phi,z)\) about the rotation axis \(\hat{\mathbf{z}}\), the effective potential acquires the centrifugal form
\[
U_{\rm cent}(r) = U(r) - \tfrac12m\omega^2 r^2,
\]
so that the radial distribution at equilibrium is
\[
\rho(r)\propto \exp\!\Bigl[-\beta\bigl(U(r)-\tfrac12m\omega^2r^2\bigr)\Bigr].
\]
This exponential enhancement at larger \(r\) is the hallmark of the outward “centrifugal” tendency in a rotating ensemble.  In contrast, the “centripetal” force \(F_{\rm centripetal}=m\omega^2r\) is not a separate potential but the real inward constraint (e.g.\ tension in a rotating rod) that enforces circular motion in an inertial frame.  In statistical mechanics it enters only insofar as it shapes the confining potential \(U(r)\), not as an additional Boltzmann factor.

Fluctuations in angular momentum satisfy
\[
\langle L_z^2\rangle - \langle L_z\rangle^2 = k_{\rm B}T \,I
\]
where \(I=mr^2\) is the moment of inertia, reflecting how thermal agitation competes with rotational ordering.  Centrifugal broadening of the radial profile scales with temperature as well as \(\omega\), whereas centripetal constraints simply truncate the accessible region of phase space at some \(r_{\max}\), independent of \(T\). In summary, centrifugal effects appear naturally in the Gibbs factor via the \(-\omega L\) coupling and manifest as an effective outward potential \(\tfrac12m\omega^2r^2\), whereas centripetal forces remain part of the underlying conservative potential enforcing boundary conditions rather than altering the statistical weight.

\subsection{Centrifugal and centripetal forces in discrete multivariate multimodal distributions}

Let \(\{x_i\}_{i=1}^M\) be the support points and \(p_i\) the probability mass at \(x_i\) of a discrete multivariate multimodal distributions.  One can introduce an energy landscape \(E_i\) so that  
\[
p_i = \frac{1}{Z}\exp\bigl(-\beta E_i\bigr)
\]  
with \(Z=\sum_j\exp(-\beta E_j)\) and \(\beta>0\) an inverse “temperature.”  Centrifugal forces in a continuous rotating ensemble correspond to a negative quadratic potential pushing mass outward; in the discrete analogue they appear as terms in \(E_i\) that decrease energy for points “far” from a center, thereby increasing \(p_i\) at the periphery.  By contrast, centripetal forces are modeled by mode‐specific wells that lower energy near each mode center \(\mu_k\).  If there are \(K\) modes, define for each \(i\)  
\[
E_i = U_{\rm base}(x_i) + \sum_{k=1}^K \frac{1}{2}\kappa_k\,\|x_i-\mu_k\|^2
\]  
where \(U_{\rm base}(x_i)\) encodes any global preference (e.g.\ a prior), \(\kappa_k>0\) is a stiffness parameter for mode \(k\), and \(\|\cdot\|\) a norm (often Euclidean).  The term \(\tfrac12\kappa_k\|x_i-\mu_k\|^2\) is a discrete centripetal potential well that “attracts” probability mass toward \(\mu_k\).  The resulting stationary distribution  
\[
p_i \propto \exp\!\Bigl[-\beta\Bigl(U_{\rm_base}(x_i)+\tfrac12\sum_k\kappa_k\|x_i-\mu_k\|^2\Bigr)\Bigr]
\]  
concentrates mass in the vicinity of each mode, with sharper peaks as \(\beta\kappa_k\) grows.  One may view the partitioning into basins of attraction by defining a graph with weights  
\[
w_{ij} = \exp\bigl(-\beta[E_i+E_j]\bigr)\!,
\]  
so that discrete “forces”  
\[
F_{i\to j} = w_{ij}(x_j-x_i)
\]  
drive mass along edges toward lower‐energy nodes.  In this framework centrifugal effects would correspond to adding negative curvature in \(U_{\rm base}\), whereas centripetal wells localize the mass.  Thermal fluctuations (finite \(\beta\)) allow hopping between wells, with transition rates satisfying a detailed balance condition  
\[
\frac{T_{i\to j}}{T_{j\to i}} = \exp\bigl(-\beta[E_j-E_i]\bigr).
\]  
Thus centripetal forces in the discrete multimodal ensemble play the role of confining potentials around modes, counteracting any centrifugal‐like dispersion encoded in the global energy landscape.

\subsection{Centrifugal and centripetal forces in continuous multivariate multimodal distributions}

A continuous multivariate multimodal ensemble may be formulated by introducing a Gibbs density
\[
p(x)=\frac{1}{Z}\exp\bigl(-\beta E(x)\bigr)
\]
with \(x\in\mathbb{R}^{D}\), inverse temperature \(\beta=1/(k_{\mathrm B}T)\), and normalization \(Z=\int_{\mathbb{R}^{D}}\exp(-\beta E(x))\,dx\).  The total energy \(E(x)\) decomposes into a baseline landscape \(U_{\mathrm{base}}(x)\), a collection of centripetal wells that confine density around each mode, and a centrifugal term that repels density from a chosen center.  If there are \(K\) modes located at \(\mu_{k}\in\mathbb{R}^{D}\) with stiffnesses \(\kappa_{k}>0\), and if the centrifugal effect is encoded by a positive‐definite matrix \(C\) with strength parameter \(\lambda>0\), one may write
\[
E(x)=U_{\mathrm{base}}(x)+\sum_{k=1}^{K}\tfrac12\kappa_{k}\|x-\mu_{k}\|^{2}\;-\;\tfrac12\lambda\,x^{\mathsf T}C\,x.
\]
Here \(\|\cdot\|\) denotes the Euclidean norm and \(x^{\mathsf T}C\,x\) is a quadratic form defining the directions of maximum repulsion.  The centripetal terms \(\tfrac12\kappa_{k}\|x-\mu_{k}\|^{2}\) create mode‐centered wells whose depths grow with \(\kappa_{k}\), thereby attracting probability density toward each \(\mu_{k}\).  In contrast the negative quadratic term \(-\tfrac12\lambda\,x^{\mathsf T}C\,x\) lowers energy for large projections of \(x\) along the eigenvectors of \(C\), pushing density outward in those directions and thus generating a centrifugal dispersion.

The resulting stationary density
\[
p(x)\propto\exp\bigl[-\beta\bigl(U_{\mathrm{base}}(x)+\tfrac12\sum_{k}\kappa_{k}\|x-\mu_{k}\|^{2}-\tfrac12\lambda\,x^{\mathsf T}C\,x\bigr)\bigr]
\]
inherits both confining and repulsive curvature.  The local force field follows
\[
F(x)=-\nabla E(x)=-\nabla U_{\mathrm{base}}(x)\;-\;\sum_{k}\kappa_{k}(x-\mu_{k})\;+\;\lambda\,C\,x,
\]
so that each centripetal term \(-\kappa_{k}(x-\mu_{k})\) draws samples toward the corresponding mode while the centrifugal contribution \(\lambda\,C\,x\) pushes them away from the global origin (or other chosen center).  Thermal fluctuations of magnitude \(T\) mediate hopping between neighbouring basins, with transition rates \(T_{x\to y}\) satisfying detailed balance
\[
\frac{T_{x\to y}}{T_{y\to x}}=\exp\bigl[-\beta\bigl(E(y)-E(x)\bigr)\bigr].
\]
As \(\beta\kappa_{k}\) grows the density sharpens within each well, whereas increasing \(\beta\lambda\) exaggerates the centrifugal thinning of density between modes, yielding an equilibrium that balances confinement and dispersion in a manner entirely analogous to rotating continuous media.

\section{Multivariate multimodal distributions of inequality and multidimensional metrics of inequality}

\subsection{Multivariate multimodal distributions of inequality}

A family of multimodal distributions of inequality can be defined from multivariate Beta distributions or from a multivariate mixture of Beta distributions. Multivariate beta distributions, such as the Dirichlet (for simplex-constrained vectors) and matrix-variate beta (for positive-definite matrices), generalize the univariate beta distribution to multiple dimensions but are typically rigid, unimodal, and have constrained support with inherent dependencies. In contrast, a multivariate mixture of beta distributions combines multiple beta components with weights, allowing flexible modeling of multimodal shapes on domains like $(0,1)^K$ and enabling customizable dependency structures via copulas, making it suitable for density estimation, clustering, and complex Bayesian hierarchical modeling where heterogeneity is present.

A mixture of Beta distributions provides a theoretically grounded and empirically robust framework for measuring inequality. Beta mixtures are naturally defined on the unit interval $[0,1]$, making them well-suited to inequality measures that must lie between 0 (complete equality) and 1 (maximum inequality). This bounded domain ensures interpretability and aligns with the scale of common inequality indices like the Gini coefficient or the Theil coefficient. Besides respecting bounded supports, a multivariate mixture of Beta distributions models multimodality and subpopulation heterogeneity, and handles skewed and heavy-tailed data—key features frequently associated with inequality in real-world phenomena, as in \citet{de2024small}.

\subsubsection{Multivariate Beta distributions and multimodality}

Let $\mathbf{x} = (x_1, x_2, \dots, x_K)$ with $x_k > 0$ and $\sum_{k=1}^{K} x_k < 1$. A multimodal extension of the multivariate Beta distribution can be expressed as a finite mixture of Dirichlet components:

\[
f(\mathbf{x}) = \sum_{m=1}^{M} \pi_m  f_D(\mathbf{x}; \boldsymbol{\alpha}^{(m)}), \quad \text{where } \sum_{m=1}^{M} \pi_m = 1, \; \pi_m \geq 0
\]

\[
f_D(\mathbf{x}; \boldsymbol{\alpha}^{(m)}) = \frac{1}{B(\boldsymbol{\alpha}^{(m)})} \prod_{k=1}^{K} x_k^{\alpha_k^{(m)} - 1}
\]

where $M$ is the number of mixture components (modes), $\pi_m$ are the mixing proportions, and $\boldsymbol{\alpha}^{(m)} = (\alpha_1^{(m)}, \dots, \alpha_K^{(m)})$ are the concentration parameters for the $m$-th component. Each component has its own mode and shape, allowing the full distribution $f(\mathbf{x})$ to exhibit multiple local maxima.

The multivariate beta normalization constant is:

\[
B(\boldsymbol{\alpha}^{(m)}) = \frac{\prod_{k=1}^{K} \Gamma(\alpha_k^{(m)})}{\Gamma\left(\sum_{k=1}^{K} \alpha_k^{(m)}\right)}
\]

This mixture model is capable of representing multimodal behavior. In the case of a $\mathbf{X} \in \mathbb{R}^{p \times p}$, i.e. a symmetric positive-definite matrix, the matrix-variate multimodal beta distribution of the second kind can be defined as a finite mixture:
\[
f(\mathbf{X}) = \sum_{m=1}^{M} \pi_m \cdot f_{B_p}(\mathbf{X}; a^{(m)}, b^{(m)}), \quad \text{where } \sum_{m=1}^{M} \pi_m = 1, \; \pi_m \geq 0
\]
Each component $f_{B_p}(\mathbf{X}; a^{(m)}, b^{(m)})$ is a matrix-variate beta density:
\[
f_{B_p}(\mathbf{X}; a, b) = \frac{|\mathbf{X}|^{a - (p+1)/2} \, |\mathbf{I} + \mathbf{X}|^{-(a + b)}}{\mathrm{B}_p(a, b)}
\]
The multivariate beta function $\mathrm{B}_p(a, b)$ is defined as:
\[
\mathrm{B}_p(a, b) = \frac{\Gamma_p(a) \Gamma_p(b)}{\Gamma_p(a + b)}, \quad a, b > \frac{p-1}{2}
\]
And the multivariate gamma function is given by:
\[
\Gamma_p(a) = \pi^{p(p-1)/4} \prod_{j=1}^{p} \Gamma\left(a - \frac{j - 1}{2} \right)
\]
This formulation allows the density $f(\mathbf{X})$ to exhibit multimodal behavior in the space of positive-definite matrices, useful for modeling heterogeneous covariance structures or uncertainty in matrix-valued latent variables.

\subsubsection{Multivariate mixture of Beta distributions and multimodality}

Let $\mathbf{x} = (x_1, x_2, \dots, x_K)^\top$ be a $K$-dimensional random vector where $x_k \in (0, 1)$. A multivariate mixture of independent $\beta$-distributions allows for flexible modeling of complex, potentially multimodal densities over $[0,1]^K$:

\[
f(\mathbf{x}) = \sum_{m=1}^{M} \pi_m \prod_{k=1}^{K} \frac{1}{\mathrm{B}(\alpha_k^{(m)}, \beta_k^{(m)})} \, x_k^{\alpha_k^{(m)} - 1} (1 - x_k)^{\beta_k^{(m)} - 1}
\]

where $M$ is the number of mixture components, $\pi_m$ are the mixing proportions with $\sum_{m=1}^{M} \pi_m = 1$, $\pi_m \geq 0$, and $\mathrm{B}(\alpha_k^{(m)}, \beta_k^{(m)})$ is the beta function defined as:

\[
\mathrm{B}(\alpha, \beta) = \frac{\Gamma(\alpha) \Gamma(\beta)}{\Gamma(\alpha + \beta)}
\]

Each mixture component represents an independent product of $K$ univariate beta distributions with its own shape parameters $(\alpha_k^{(m)}, \beta_k^{(m)})$ per variable $x_k$. The overall mixture $f(\mathbf{x})$ is capable of exhibiting multimodality, with modes determined by the local maxima of each mixture component and their interactions in the product space. This type of mixture of $\mathcal{B}_d$ distributions---often used in Bayesian nonparametrics, density estimation, and latent class models---are a flexible alternative to the Dirichlet distribution when dependence across components is not enforced via the simplex constraint but rather through the shape of marginal distributions and the combination of modes across dimensions. While this formulation covers independent marginals, it allows multimodal joint behavior via the mixture. For dependent beta marginals or constrained supports (e.g., the simplex), a Dirichlet mixture model is more appropriate.

\subsection{Multidimensional metrics of inequality based on multivariate multimodal distributions}

The multivariate Beta distribution, such as the Dirichlet or matrix-variate beta, is generally not well-suited for modeling multimodal data. These distributions are unimodal when parameters are greater than one and may exhibit skewness or concentration effects, but they lack the flexibility to capture multiple modes across the support. Their structure is constrained, either by the simplex geometry in the case of the Dirichlet, or by matrix properties in the case of the matrix-variate beta, making them inadequate for representing heterogeneous or clustered data.

In contrast, a multivariate mixture of beta distributions is highly suitable for multimodal modeling. By combining multiple beta components, each with its own shape parameters and weight, the mixture can represent complex marginal behaviors, capture distinct subpopulations, and approximate arbitrary density shapes over bounded domains. When extended with copula-based dependence structures, these mixtures can model both multimodality and inter-variable dependencies, offering a powerful and flexible tool for density estimation, clustering, and Bayesian hierarchical modeling in settings where the data exhibit multiple modes or latent heterogeneity.

An inequality metric $\mathcal{I}$ an be defined for a multivariate mixture of beta distributions as a composite functional that captures three key structural dimensions of inequality: (1) the number of modes, reflecting population stratification; (2) the concentration of probability mass within each mode, reflecting intra-group inequality or homogeneity; and (3) the geometric separation between modes, reflecting inter-group disparity. Given a fitted mixture model:

\[
f(\mathbf{x}) = \sum_{m=1}^{M} \pi_m f_m(\mathbf{x})
\]

where each $f_m(\mathbf{x})$ is a component density (e.g., product of beta marginals with copula dependencies), and $\pi_m$ the mixing weights, we propose the following inequality index:

\[
\mathcal{I} = \underbrace{H(\boldsymbol{\pi})}_{\text{modal entropy}} \left( \sum_{m=1}^{M} \pi_m  \kappa(f_m) \right) \mathbb{D}
\]

Here, $H(\boldsymbol{\pi}) = -\sum_{m=1}^{M} \pi_m \log \pi_m$ quantifies the \textbf{modal entropy}, which increases with the number of effective modes and their balance, encoding multimodality. The term $\kappa(f_m)$ measures the \textbf{concentration} (e.g., via inverse differential entropy or peakiness) of the $m$-th mode, reflecting how tightly probability is clustered around each mode. Finally, $\mathbb{D}$ denotes the \textbf{average pairwise distance} between modal centers $\boldsymbol{\mu}_m$, weighted by $\pi_m \pi_{m'}$, such as:

\[
\mathbb{D} = \sum_{m < m'} \pi_m \pi_{m'} d(\boldsymbol{\mu}_m, \boldsymbol{\mu}_{m'})
\]

where $d(\cdot, \cdot)$ is a suitable distance metric in $\mathbb{R}^K$ (e.g., Mahalanobis or Wasserstein). This formulation captures inequality as a function of how many distinct regions of high probability exist, how sharply each is defined, and how far apart they are—emphasizing the coexistence of fragmentation and disparity in distributional geometry. The metric $\mathcal{I}$ is particularly relevant in socioeconomic, ecological, and epidemiological applications where inequality arises not only from spread or skewness, but from structured, clustered heterogeneity within bounded or compositional variables.

\section{Monte Carlo Simulation Experiments}

Define a multivariate mixture of Beta distributions in two dimensions, where each observation $\mathbf{x} = (x_1, x_2)$ lies in the unit square $[0,1]^2$. The mixture model is given by
$$
f(\mathbf{x}) = \sum_{m=1}^{M} \pi_m f_m(x_1, x_2)
$$
where each component $f_m(x_1, x_2)$ is the product of two independent Beta distributions:
$$
f_m(x_1, x_2) = \mathcal{B}(x_1 \mid \alpha_1^{(m)}, \beta_1^{(m)}) \times \mathcal{B}(x_2 \mid \alpha_2^{(m)}, \beta_2^{(m)})
$$
In the Monte Carlo (MC) simulation experiments, samples were generated from this mixture using random variates from the Beta distribution. In each MC experiment, the modal entropy quantifies the degree of multimodality and balance in the mixture weights $\boldsymbol{\pi} = (\pi_1, \dots, \pi_M)$, and is defined as
$$
H(\boldsymbol{\pi}) = -\sum_{m=1}^{M} \pi_m \log \pi_m
$$
Higher modal entropy corresponds to more balanced modes and greater structural inequality, while lower entropy reflects dominance by a single mode.

The concentration of each mode is measured by the inverse of the Shannon entropy of a discretized 2D histogram of the sampled distribution. For each component $m$, the density is estimated over a grid, and its concentration is defined as

$$
\kappa(f_m) = \frac{1}{H_m}, \quad H_m = -\sum_i p_i \log p_i
$$

where $p_i$ are the probabilities of the histogram bins. A sharper, more peaked component results in higher $\kappa(f_m)$. The total concentration term is aggregated as a weighted average:

$$
\sum_{m=1}^{M} \pi_m  \kappa(f_m)
$$

This term reflects the degree of internal uniformity or homogeneity within each mode. The third term captures the average pairwise Euclidean distance between mode centers, defined as

$$
\boldsymbol{\mu}_m = \left( \frac{\alpha_1^{(m)}}{\alpha_1^{(m)} + \beta_1^{(m)}}, \frac{\alpha_2^{(m)}}{\alpha_2^{(m)} + \beta_2^{(m)}} \right)
$$

and the distance metric is given by

$$
\mathbb{D} = \sum_{m < m'} \pi_m \pi_{m'} d(\boldsymbol{\mu}_m, \boldsymbol{\mu}_{m'})
$$

where $d(\cdot, \cdot)$ is the Euclidean distance in $\mathbb{R}^2$. This term captures the degree of separation or polarization between modes. The final inequality metric combines these three dimensions multiplicatively:
$$
\mathcal{I} = H(\boldsymbol{\pi}) \left( \sum_{m=1}^{M} \pi_m \kappa(f_m) \right) \mathbb{D}
$$
This metric increases with the number and balance of modes, the concentration of probability within each mode, and the geometric separation between modes. It provides an interpretable measure of polarizing inequality over bounded domains, capturing stratification, segregation, and disparity in a unified probabilistic framework.

\begin{figure}[ht!]
    \centering
    \includegraphics[width=\linewidth]{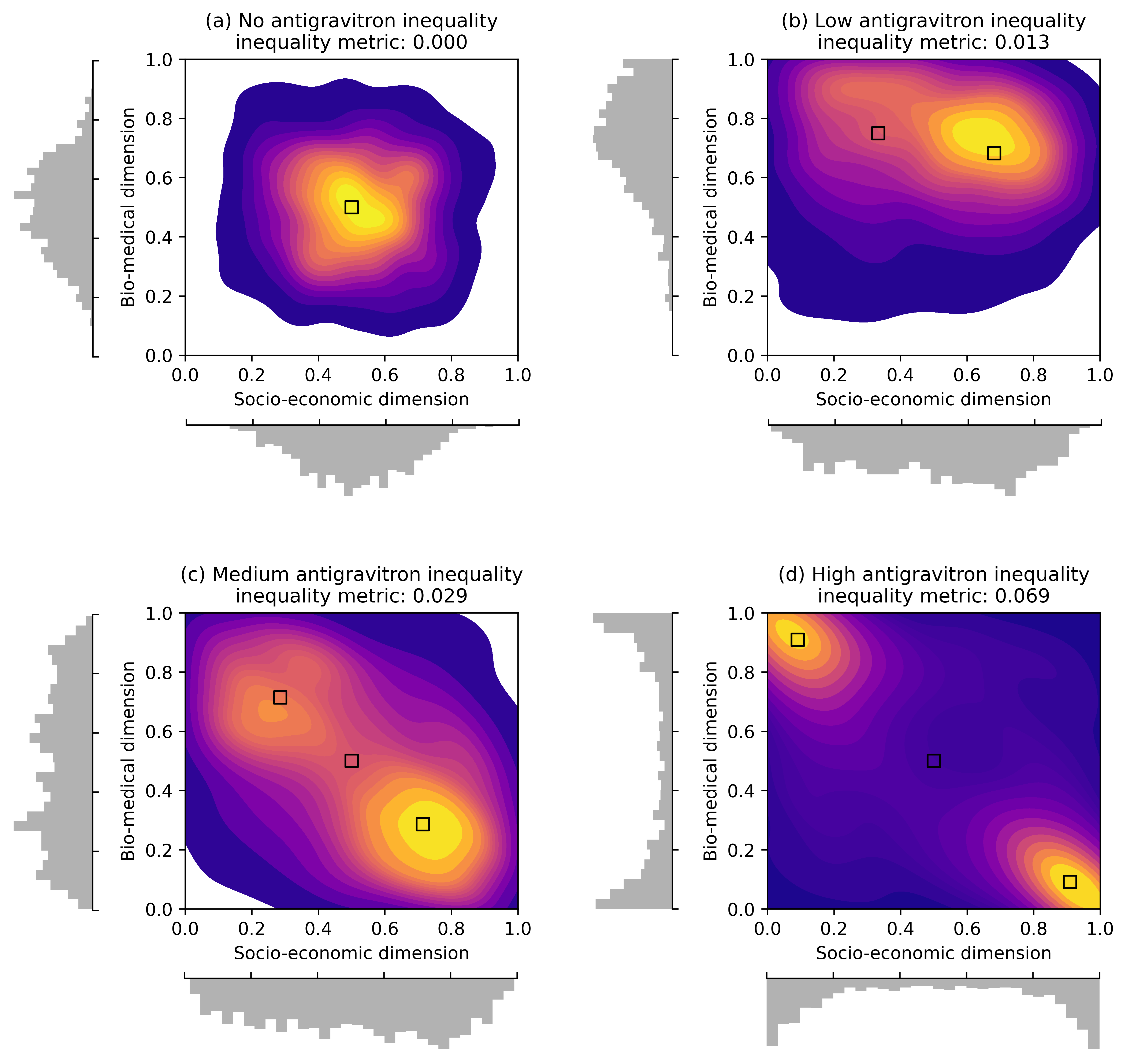}
    \caption{Joint density plots and marginal histograms for four simulated scenarios of AntiGravitron inequality. Each subplot depicts increasing levels of multimodality and density stratification across socio-economic and bio-medical dimensions. Black squares represent mode centers. The inequality metric increases with the number of modes, the sharpness of local concentration, and the spatial distance between modal centers, illustrating how probabilistic inequality can be geometrically quantified in a bounded multivariate framework inspired by (anti)gravitron forces from statistical physics.}
    \label{fig:MCsim}
\end{figure}

Figure \ref{fig:MCsim} shows the results of Monte Carlo simulation from four scenarios of inequality. Each scenario represents distinct levels of AntiGravitron inequality forces that pulls probability density into distinct modes on the joint space defined by a socio-economic and bio-medical dimensions.

In the first subplot (\ref{fig:MCsim}a), the distribution is unimodal and symmetric across both axes, with high entropy and minimal concentration or separation. This results in an inequality metric equal to zero, representing a completely homogeneous scenario where individuals cluster around a central equilibrium of socio-economic and bio-medical attributes.

The second subplot (\ref{fig:MCsim}b) introduces a low level of inequality by splitting the density into two closely located modes. Although the entropy is slightly lower and the concentration per mode is higher, the spatial distance between modes is minimal. The resulting inequality metric is small but positive, signaling early polarization.

The third subplot (\ref{fig:MCsim}c) features three moderately polarized and spatially dispersed modes. The distribution begins to exhibit noticeable multimodality, with populations clustering around different socio-biomedical poles. The polarizing inequality metric increases accordingly, capturing this emerging complexity and polarization.

In the final subplot (\ref{fig:MCsim}d), the distribution becomes highly polarized. Three sharp, distant modes concentrate density in extreme corners of the space, representing clear group separation in both socio-economic and biomedical dimensions. While the entropy is lower, the combination of strong polarization and large pairwise distances between the concentrated poles results in a high inequality metric.

Together, these scenarios illustrate how polarizing inequality can emerge from the interaction between modal structure (number of modes), local polarization (concentration and sharpness of clusters), and spatial dispersion (distance between centers). The resulting polarizing inequality metric is an interpretable, bounded (between 0 and 1), and geometrically valid measure of multivariate inequality, especially useful for modeling real-world disparities in socio-biomedical data.

\section{Empirical application to inequalities in US household income by ethnicity and geographical region}

The AntiGravitron polarizing inequality metric was empirically applied to data on median household income by ethnicity and geographic subregion within the Greater Philadelphia area of the United States. The source of the data is the Delaware Valley Regional Planning Commission (DVRPC), the federally designated Metropolitan Planning Organization (MPO) for the region, established through an interstate compact between the Commonwealth of Pennsylvania and the State of New Jersey. 

The DVRPC dataset contains variables such as survey year, geographic subregion, ethnic group, median household income disaggregated by region and ethnicity, and the income gap relative to the White (non-Hispanic) population benchmark. Previous research \citep{ho2022ethnic, sze2020ethnicity, magesh2021disparities} found robust links between ethnicity and biomedical inequality. For instance, individuals identifying as Black or Asian face higher risks of cardiovascular disease compared to their White counterparts \citep{ho2022ethnic}, and were disproportionately affected by infections during the COVID-19 pandemic \citep{sze2020ethnicity}. These disparities were further exacerbated by socioeconomic inequality and differential quality of clinical care, contributing to elevated COVID-19 mortality rates among ethnic minority groups \citep{magesh2021disparities}. Geographic patterns of socio-economic inequality reinforce these existing disparities. Economic segregation---the spatial separation of populations by income---has been shown to shape individuals’ lived experiences of inequality \citep{davidai2024economic}, with geo-clustering strongly associated with socio-economic deprivation \citep{shin2019geo}. Racialized economic segregation---the interplay of spatial, social, and income polarization---has also been directly linked to adverse health outcomes, including increased morbidity and mortality \citep{xu2024two}.

\begin{figure}[ht!]
    \centering
    \includegraphics[width=\linewidth]{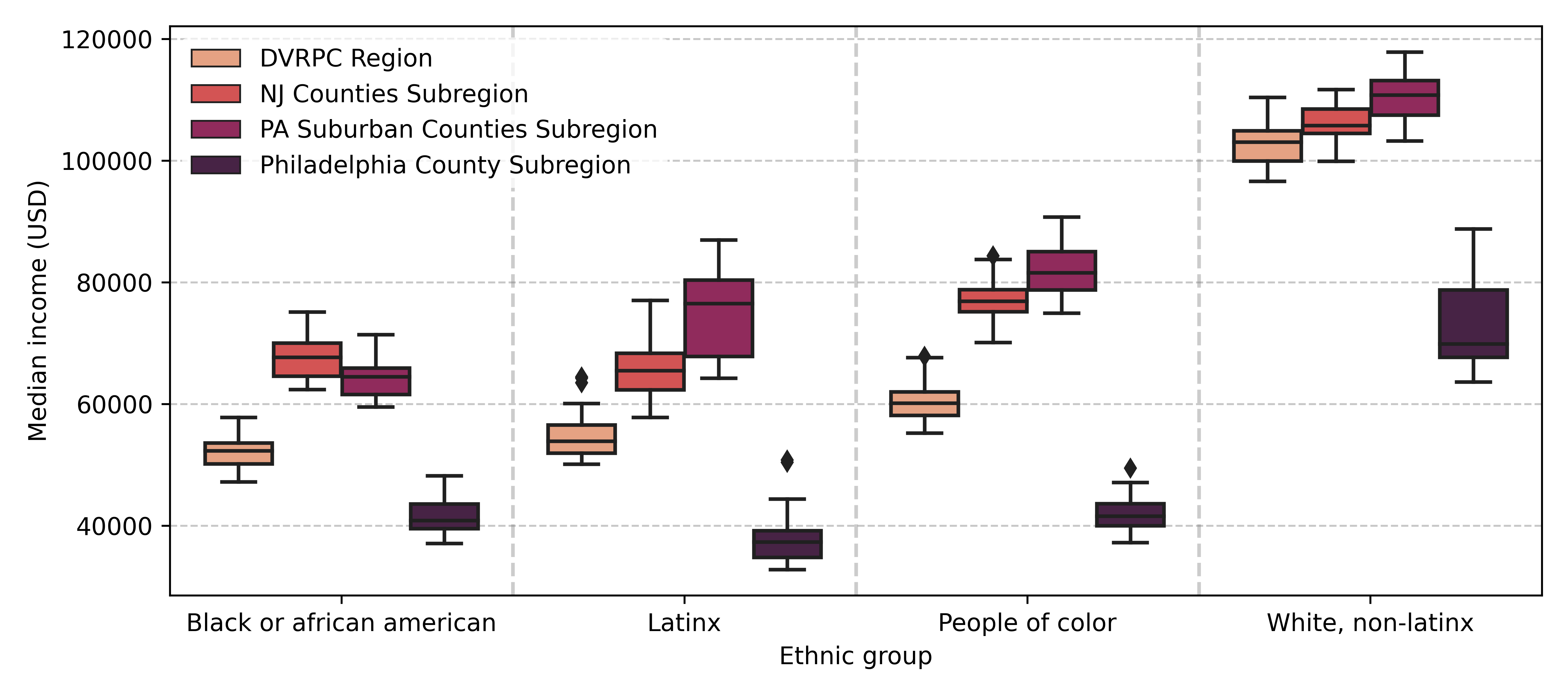}
    \caption{Median household income in US dollars, by race and ethnicity in the Greater Philadelphia region. The boxplot shows the distribution of median household incomes across ethnic groups, disaggregated by geography within the Delaware Valley Regional Planning Commission (DVRPC) region.}
    \label{fig:boxplot01}
\end{figure}

\begin{figure}[ht!]
    \centering
    \includegraphics[width=\linewidth]{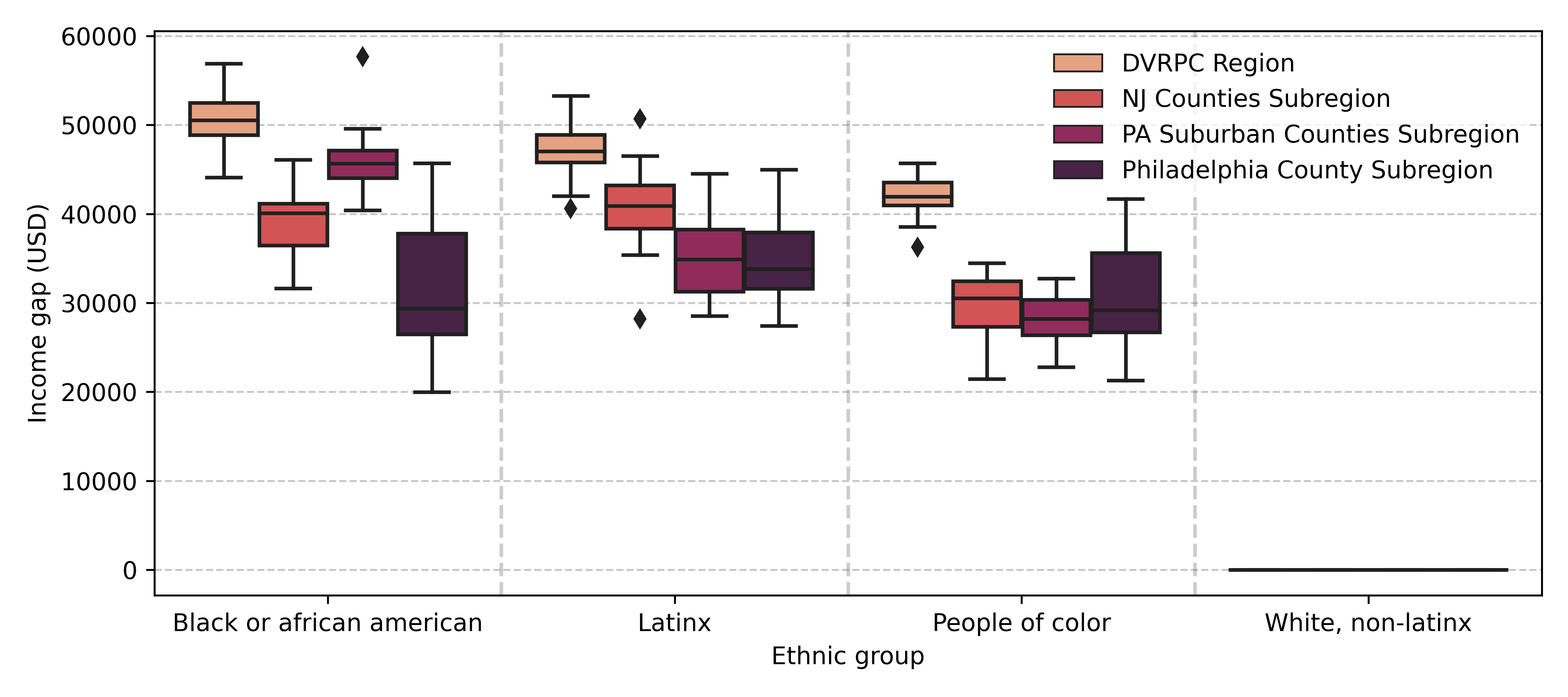}
    \caption{Income gap from white non-latinx households, by race and ethnicity in the Greater Philadelphia region. The boxplot shows the distribution of the income gap in US dollars across ethnic groups, disaggregated by geography within the Delaware Valley Regional Planning Commission (DVRPC) region.}
    \label{fig:boxplot02}
\end{figure}

Figures \ref{fig:boxplot01} and \ref{fig:boxplot02} visualize the disparities in household income across ethnic groups and geographic subregions within the Greater Philadelphia area, which were calculated with the data from the DVRPC. Figure \ref{fig:boxplot01} illustrates the distribution of median income for four ethnic groups (Black or African American, Latinx, People of Color, and White non-Latinx) across geographic subregions (the DVRPC, NJ Counties, PA Suburban Counties, and the Philadelphia County subregion). The White non-Latinx group consistently exhibits the highest median incomes, especially in the PA Suburban Counties, with values often exceeding USD 100000. Black and Latinx groups show markedly lower median incomes across all regions, particularly in the Philadelphia County Subregion, where the median household income is below USD 50000. The PA Suburban Counties Subregion tends to have the widest range of incomes within each ethnic group, suggesting higher intra-group variation (that is, heterogeneity). The household in the Philadelphia County have the lowest income medians and also the most compressed distributions, reflecting concentrated disadvantage and polarization in those regions.

Figure \ref{fig:boxplot02} shows the absolute household income gap between each ethnic group and the White non-Latinx baseline, across the same subregions. Black or African American and Latinx groups have the largest income gaps, often exceeding USD 50000, especially in the PA Suburban Counties regions. The Philadelphia County Subregion shows smaller gaps on average for Black and Latinx populations, but not due to increased incomes for these groups, but rather due to the significantly lower median income of the White group in this subregion (as seen in Figure \ref{fig:boxplot01}). People of Color (aggregated) show consistent income disadvantages, though slightly less pronounced than the more specific categories. The boxplot highlights interquartile variability, indicating that the gap is not uniform within group-region combinations.

These figures jointly expose spatial and ethnic inequality in economic outcomes. The household income disparities are systemic and regionally differentiated, with ethnicity and geography interacting in ways that exacerbate or slightly mitigate inequality. They also provide a compelling visualization of intersectional disadvantage and polarization, where groups are multidimensional marginalized based on both their ethnic identity and their location.

The estimation results of the AntiGravitron polarizing-inequality metric applied to the US data (Table \ref{tab:resineqUS}) show spatial and ethnic heterogeneities in distributional concentration and modal divergence across the DVRPC region and its subregions. The Philadelphia County is the region with the highest measured of polarizing inequality across all ethnic categories, particularly among Black or African American individuals ($1.67 \times 10^{-2}$), followed by People of Color ($1.43 \times 10^{-2}$), and Latinx population groups ($1.19 \times 10^{-2}$). This finding is expected and hosw the ability of the proposed metric to capture polarizing inequality, because the  Philadelphia County households have the lowest median income, compared to households in other regions (Figure \ref{fig:boxplot01}). At ethnic level, the high polarizing inequality suggests a confluence of economic marginalization and spatial concentration that amplifies distributional differences compared to White, non-latinx population groups, which is indicative of potential systemic exclusion or urban economic polarization mechanisms \citep{chetty2018race, powell2008structural}.

\renewcommand{\arraystretch}{1.1}
\begin{table}[ht!]
    \centering
    \caption{Estimation results of AntiGravitron inequality}
    \label{tab:resineqUS}
    \medskip
\begin{tabular}{llc}
   \toprule
        \textbf{region} & \textbf{ethnicity} & \textbf{inequality} \\ \midrule
            & Black or african american & 7.29 $\times 10^{-3}$ \\
        DVRPC & Latinx & 7.00 $\times 10^{-3}$ \\
            & People of color & 5.71 $\times 10^{-3}$ \\ \hline
            & Black or african american & 1.02 $\times 10^{-2}$ \\ 
        NJ counties & Latinx & 9.29 $\times 10^{-3}$ \\ 
            & People of color & 8.73 $\times 10^{-3}$ \\ \hline
            & Latinx & 1.55 $\times 10^{-2}$ \\ 
        PA suburban counties & People of color & 8.19 $\times 10^{-3}$ \\ 
            & Black or african american & 7.54 $\times 10^{-3}$ \\ \hline
            & Black or african american & 1.67 $\times 10^{-2}$ \\ 
        Philadelphia county & People of color & 1.43 $\times 10^{-2}$ \\ 
            & Latinx & 1.19 $\times 10^{-2}$ \\ \bottomrule
    \end{tabular}
\end{table}

In contrast, the broader DVRPC region exhibits comparatively lower levels of polarizing inequality, with the Black or African American group again experiencing the highest inequality ($7.29 \times 10^{-3}$), though the metric is roughly half of that observed in Philadelphia. This contrast underscores a regionally contingent expression of inequality, wherein urban cores exhibit intensified disparity compared to their suburban or regional aggregates \citep{reardon2004patterns}. New Jersey Counties and PA Suburban Counties display intermediate patterns, with Latinx and Black populations experiencing elevated polarizing inequality. The highest value in these subregions is observed among Latinx individuals living in the PA suburbs ($1.55 \times 10^{-2}$), pointing to potentially distinct suburban stratification mechanisms affecting this group \citep{bailey2017structural}.

Overall, the empirical results highlight that the AntiGravitron metric of polarizing inequality properly synthesizes distributional complexity by integrating entropy, concentration, and modal separation. These results substantiate the hypothesis that income disparities, when benchmarked against White (non-Hispanic) populations, are neither uniform across geographies nor ethnically invariant. Rather, they exhibit nuanced structural gradients of polarizing inequality related to localized socio-economic and historical-political determinants \citep{massey2009origins}. From a policy standpoint, this result underlines the necessity of geographically differentiated and ethnically responsive interventions to address polarizing inequality in income distribution.

\FloatBarrier

\section{Discussion}

The results from both the Monte Carlo simulations and empirical analysis validate the theoretical intuition that inequality emerges from the interplay of fragmentation, concentration, and spatial polarization. Unlike traditional scalar measures that collapse distributional complexity into a single dimension, the AntiGravitron metric of polarizing inequality captures the structural geometry of inequality in its full multidimensional form. Like \citet{chatterjee2007kinetic}, who identified cluster separation as analogous to phase fragmentation in kinetic-exchange dynamics, we  empirically model wealth exchange akin to gas kinetics, underpinning the mixture-of-attractors of the AntiGravitron.

Simulation results illustrate that even modest bifurcations in modal structure---when amplified by sharp intra-modal concentration and sufficient inter-modal distance---yield disproportionately high inequality polarization. This aligns with known phenomena in socio-economic systems where small policy shifts or network reconfigurations lead to cascading inequality amplification and social stratification into poles. In the empirical application, the Philadelphia County, characterized by intense demographic segmentation and economic polarization, exhibits the highest inequality values. These findings are consistent with critical urban theory---which emphasizes how power, inequality, and capitalist processes structure urban space \citep{rossi2018critical}---and ecological models of segregation---which conceptualize segregation as a resilient, adaptive system maintained through institutional and environmental mechanisms \citep{pickett2023resilience}---, reinforcing the idea that inequality is not merely a statistical irregularity but a product of structural self-organization.

Furthermore, the entropy-concentration-distance formulation of the proposed polarizing inequality metric aligns with foundational principles in both information theory and thermodynamics. Specifically, Shannon’s information entropy exhibits a formal mathematical equivalence to thermodynamic entropy as established by Gibbs and Boltzmann \citep{shannon1948mathematical, gibbs1902elementary}. Moreover, the structure of the Helmholtz free energy function \citep{helmholtz1882thermodynamics} conceptually parallels the AntiGravitron trade-off between structural concentration (internal energy) and distributional entropy (disorder). The flexibility of Beta mixture models in capturing the dynamic structure of the AntiGravitron framework makes the proposed polarizing inequality metric particularly robust to the bounded, skewed, and multimodal characteristics commonly observed in real-world inequality data, thus addressing critical limitations of traditional Gaussian-based approaches. Moreover, the capacity to model the "inequality of inequality" across polarized subpopulations introduces a second-order diagnostic capability, enabling the detection of latent structural disparities and the identification of critical bifurcation zones within socio-economic and bio-medical systems.

\section{Conclusion}

This paper proposed a novel multidimensional metric of polarizing inequality grounded in statistical physics and inspired by the dynamics of a conceptual AntiGravitron that illustrates how inequality arises due societal centripetal and centrifugal polarizing forces, potentially with a Coriolis effect. By integrating entropy, modal concentration, and mass separation into a unified probabilistic framework, the polarizing inequality metric proposed in this study goes beyond traditional scalar indices to model inequality as an emergent, bifurcating, and spatially heterogeneous process.

The multivariate Beta mixture framework used to calculate the AntiGravitron inequality is a statistically coherent approach for modeling bounded and heavy-tailed distributions and is also metaphorically resonant with theories of social stratification, systemic feedback, and institutional magnetodynamics. The Monte Carlo simulations and the empirical application to income disparities in the Greater Philadelphia area show the metric’s practical relevance and its capacity to expose spatially polarizing inequalities that traditional measures obscure.

Future research can extend the polarizing inequality approach to dynamic settings through Bayesian methods, exploring how interventions reshape the configuration of modal basins and what conditions lead to either the attenuation and amplification of polarization. By embedding inequality measurement within the language of energy landscapes and rotating field systems, this study opens a conceptual bridge between the social and physical sciences, thus allowing inequality to be theorized not as a normative deviation from equilibrium, but as a structural attractor in complex adaptive systems.

\printbibliography

\end{document}